\documentclass[prd,a4paper,nofootinbib,showpacs]{revtex4}
\usepackage[a4paper, hdivide={2.5cm,,2.5cm}, vdivide={3cm,,3.5cm}]{geometry}

\usepackage{amstext,amssymb}
\usepackage[intlimits]{amsmath}
\usepackage[english]{babel}
\usepackage[ansinew]{inputenc}
\usepackage{graphicx}
\usepackage{slashed} 
\usepackage{dsfont} 
\usepackage{units}
\usepackage{hyperref}

\hypersetup{
    pdftitle={Seesaw parametrization for n right-handed neutrinos},    
    pdfauthor={Julian Heeck}     
}

\begin{document}

\let \Lold \L
\def \L {\mathcal{L}} 
\let \epsilonold \epsilon
\def \epsilon {\varepsilon} 
\let \arrowvec \vec
\def \vec#1{{\boldsymbol{#1}}}
\newcommand{\del}{\partial}
\newcommand{\dd}{\mathrm{d}}
\newcommand{\matrixx}[1]{\begin{pmatrix} #1 \end{pmatrix}} 
\newcommand{\tr}{\mathrm{tr}}
\newcommand{\hc}{\mathrm{h.c.}}
\newcommand{\const}{\mathrm{const.}}
\newcommand{\BR}{\mathrm{BR}}
\newcommand{\re}{\mathrm{Re}\,}
\newcommand{\im}{\mathrm{Im}\,}
\newcommand{\diag}{\mathrm{diag}}
\def \eins {\mathds{1}} 
\def \M {\mathcal{M}} 

\renewcommand{\thefootnote}{\fnsymbol{footnote}} 

\title{Seesaw parametrization for \texorpdfstring{$n$}{n} right-handed neutrinos}

\author{Julian \surname{Heeck}}
\email{julian.heeck@mpi-hd.mpg.de}
\affiliation{Max--Planck--Institut f\"ur Kernphysik, Saupfercheckweg 1, 69117 Heidelberg, Germany}



\begin{abstract}
Introducing $n$ right-handed neutrinos to the Standard Model yields, in general, massive active neutrinos. We give explicit parametrizations for the involved mixing and coupling matrices in terms of physical parameters for both the top-down and the bottom-up approach for arbitrary $n$. Bounds on the complex mixing angles in the bottom-up approach from perturbativity of the Yukawa couplings to charged lepton flavor violation are discussed. As a novel illustration of possible effects from $n\neq 3$, we extend the neutrino anarchy framework to arbitrary $n$; we show that while the anarchic mixing angles are insensitive to the number of singlets, the observed ratios of neutrino masses prefer small $n$ for the simplest linear measure.
\end{abstract}

\maketitle


\section{Introduction}

The (type-I) seesaw mechanism~\cite{seesaw} is arguably the simplest and best motivated framework not only to give neutrinos mass but also to explain the smallness of said masses compared to the electroweak scale. The necessary right-handed neutrino partners are assumed to be heavy to suppress the active neutrino masses, which also allows for baryogenesis through leptogenesis~\cite{leptogenesis} via the decay of the heavy states, thus explaining the baryon asymmetry of the Universe. While initially motivated in the context of grand unified theories, the seesaw mechanism has since been applied in its own right. The number of right-handed neutrinos, transforming as singlets of the Standard Model (SM) gauge group, is then no longer constrained by anomaly cancellations or the need to fill up an irreducible representation, but rather a free parameter. At least two right-handed neutrinos are necessary to reproduce the observed neutrino mass-squared differences, as well as leptogenesis~\cite{twosinglets}. There is however no upper bound, and models with $\mathcal{O}(10^2$--$10^3)$ singlets have been studied in the context of leptogenesis~\cite{Eisele:2007ws}, lepton flavor violation (LFV)~\cite{Ellis:2007wz} and as a way to explain the large mixing angles of the active neutrinos~\cite{Feldstein:2011ck}. Like for most things in life, there is also a motivation from string theory~\cite{strings}. The number of singlets is formally infinite in extra-dimensional theories, as right-handed neutrinos are not restricted to the SM-brane and will thus lead to a tower of Kaluza-Klein excitations~\cite{kaluzaklein}.

Still lacking, however, is a proper parametrization of the arising mixing and coupling matrices in terms of physical quantities, as needed for example for efficient parameter scans (as the number of redundant parameters explodes for large $n$). The goal of this work is to provide exactly these parametrizations for arbitrary $n$, both for the most general case (top-down approach, to some degree discussed in Ref.~\cite{arbitraryn}) and the seesaw limit (bottom-up).

The rest of this paper is organized as follows: we first fix our notation in Sec.~\ref{sec:framework}, discuss the top-down parametrization in Sec.~\ref{sec:top-down} and the bottom-up parametrization in Sec.~\ref{sec:bottom-up}, also deriving constraints on the involved parameters from perturbativity and LFV. To illustrate possible effects of $n$ singlets we discuss basis independence (neutrino anarchy) in this framework and present the resulting distributions for the neutrino masses in Sec.~\ref{sec:anarchy}. App.~\ref{sec:thegroupUN} gives a brief introduction to the group $U(N)$ and various representations of its elements, as needed for our parametrizations. App.~\ref{sec:measures} collects and extends basis independent measures for miscellaneous matrices necessary for the discussion of anarchy. Finally, App.~\ref{sec:nequal2} is devoted to the somewhat special case $n=2$, which would interrupt the flow if included in the main text, which will only cover $n\geq 3$.

\section{Framework and Notation}
\label{sec:framework}

\addtocounter{footnote}{2}

Introducing $n$ SM singlet right-handed neutrinos $N_j$ to the Standard Model modifies the Lagrangian by the following terms:
\begin{align}
 \L \ = \ \L_\mathrm{SM} + i \overline{N}_j \slashed{\del} N_j -\left( \overline{N}_j  \left( \vec{Y}_\nu \right)_{j\, \alpha} \tilde{H}^\dagger L_\alpha + \frac{1}{2} \overline{N}_j \left( \M_R \right)_{jk} N_k^c + \hc \right),
\end{align}
where a sum over $\alpha = e,\mu,\tau$ (sometimes $1,2,3$ in the following) and $j,k = 1,2,\dots,n$ is understood and will in the following often be denoted in a vector notation (e.g.~as $\overline{N} \M_R N^c$). Without loss of generality we work in a basis where the charged lepton mass matrix is diagonal, which can be accomplished by unitary transformations of the lepton fields. $\M_R$ can be diagonalized by an $n\times n$ unitary matrix $V$ like $V^T \M_R V = \diag$. This merely redefines $\vec{Y}_\nu$, so from now on we will work in a basis where $\M_R$ is diagonal, with positive real entries $M_i$.\footnote{This is not only convenient but also necessary to eliminate unphysical parameters. To this effect we note that a diagonal $\M_R$ can store $\mathcal{O}(n)$ parameters (together with $\vec{Y}_\nu$), while a non-diagonal $\M_R$ has $\mathcal{O}(n^2)$ independent entries.} After the Higgs doublet $H$ acquires its vacuum expectation value $v$, the Dirac mass matrix for the neutrinos $m_D = v \vec{Y}_\nu$ is generated, leading to a $(3+n)\times (3+n)$ Majorana mass matrix for the neutral fermions $(\nu_L, N^c)^T$:\footnote{Including a type-II seesaw contribution would fill the upper-left zero matrix. For explicit parametrizations of the unitary matrix that diagonalizes $\M_\mathrm{full}$ (for $n=3$) see e.g.~Refs.~\cite{Xing,blennow}.}
\begin{align}
 \M_\mathrm{full} = \matrixx{ 0 & m_D^T \\ m_D & \M_R} .
\end{align}
Assuming $\M_R \gg m_D$ gives the low-energy neutrino mass matrix for the flavor neutrinos $\nu_f \simeq \nu_L - m_D^\dagger \M_R^{-1} N^c$ of seesaw form:
\begin{align}
 \M_\nu \simeq -m_D^T \M_R^{-1} m_D = -v^2 \, \vec{Y}_\nu^T \M_R^{-1} \vec{Y}_\nu \,.
\label{eq:seesawI}
\end{align}
Diagonalization of $\M_\nu$ can be performed in the following way
\begin{align}
 \M_\nu = U_\mathrm{PMNS}^*\, \diag (m_1, m_2, m_3) \,U_\mathrm{PMNS}^\dagger \equiv  U_\mathrm{PMNS}^*\, d_m \,U_\mathrm{PMNS}^\dagger \,,
\end{align}
with the unitary Pontecorvo-Maki-Nakagawa-Sakata matrix (PMNS matrix) in standard parametrization
\begin{align}
	U_\mathrm{PMNS} =
P' \,\matrixx{c_{12} c_{13} & s_{12} c_{13} & s_{13} e^{-i\delta}\\
	-c_{23} s_{12}- s_{23} s_{13} c_{12} e^{i\delta} & c_{23} c_{12}- s_{23} s_{13} s_{12} e^{i\delta} & s_{23} c_{13}\\
	s_{23}s_{12}- c_{23} s_{13} c_{12} e^{i\delta} & -s_{23} c_{12}- c_{23} s_{13} s_{12} e^{i\delta} & c_{23} c_{13}}  \,P \,,
\label{eq:pmns}
\end{align}
and the Majorana phase matrix $P = \diag (1, e^{i \alpha/2}, e^{i\beta/2})$. Here we used the abbreviations $s_{ij} \equiv \sin \theta_{ij}$ and $c_{ij} \equiv \cos \theta_{ij}$ for the three mixing angles. The phase matrix $P' = \diag (e^{i a/2}, e^{i b/2}, e^{i c/2})$ can be absorbed by the lepton fields and is hence unphysical. The neutrino mass eigenstates are then $\nu_m = U_\mathrm{PMNS}^\dagger \nu_f$.

At low energies the $9$ complex entries of the symmetric Majorana matrix $\M_\nu$ decompose into the three eigenvalues $m_i$ (neutrinos masses), three mixing angles $\theta_{23}$, $\theta_{12}$, $\theta_{13}$ (atmospheric, solar and reactor angle) and three CP violating phases $\delta,\alpha,\beta$, the latter two being unobservable in neutrino oscillations (but in principle testable in $0\nu \beta\beta$ experiments~\cite{betadecay}).

\section{Top-down parametrization}
\label{sec:top-down}

All the information about neutrino mixing is encoded in $\vec{Y}_\nu$, the only non-diagonal matrix in the lepton sector. For the general complex $n\times 3$ matrix $\vec{Y}_\nu$, we can write down a singular value decomposition
\begin{align}
 \vec{Y}_\nu = V_R \matrixx{\diag (y_1, y_2, y_3)\\ \arrowvec{0}_3\\\vdots\\\arrowvec{0}_3} V_L^\dagger \equiv  V_R D_Y V_L^\dagger\,,
\label{eq:VRparam2}
\end{align}
with positive singular values $y_i$. Out of the $6 n$ real parameters in $\vec{Y}_\nu$, three phases can be absorbed in the lepton fields, so only $6 n -3$ are physical. We will confirm this with an explicit parametrization of the unitary matrices $V_{R,L}$ below. For the $3\times 3$ matrix $V_L$, we take the PMNS parametrization from Eq.~\eqref{eq:pmns}, while the $n\times n$ matrix $V_R$ can be written as
\begin{align}
 V_R = \left(\prod_{i=1}^n \prod_{j=i+1}^n \Omega_{ij} (\alpha_{ij},\xi_{ij}) \right) \times \diag (e^{i \phi_1},\dots,e^{i \phi_n})\,.
\label{eq:vr}
\end{align}
Here, $\Omega_{ij}  (\alpha_{ij},\xi_{ij}) $ denotes matrices of the form
\begin{align}
 \Omega_{ij}  (\alpha_{ij},\xi_{ij})  \equiv \matrixx{ 1 & &  & & & \\ & \ddots & & & &  & \\ & & \cos (\alpha_{ij}) & &  \sin (\alpha_{ij}) e^{i \xi_{ij}} &  & \\ & & & \ddots & &  & \\ & & - \sin (\alpha_{ij}) e^{-i \xi_{ij}} & & \cos (\alpha_{ij}) &  & \\ & & & & &\ddots  & \\ & & & & & & 1\\} ,
\label{eq:complexrotation2}
\end{align}
where $\sin (\alpha_{ij}) e^{i \xi_{ij}}$ sits at the $i$-th row and $j$-th column.
See App.~\ref{sec:thegroupUN} for a derivation and a proof for the validity of this form of $V_R$. It is easy to check that $V_R$ ($V_L$) has $n^2$ ($3^2$) real parameters, $n (n+1)/2$ ($6$) of which are phases. From the form of $\vec{Y}_\nu$ in Eq.~\eqref{eq:VRparam2}, it is clear that the angles $\phi_j$ either act on zeros in $D_Y$ or can be absorbed by the phase matrix $P_L^\dagger$ in $V_L^\dagger$. An overall phase and ${P_L'}^\dagger$ can be absorbed by the lepton fields, so $V_L$ contains only three angles and three phases. The ordering of the $\Omega_{ij}$ in $V_R$ is of course arbitrary, which allows us to move all the rotations that act on the zeros in $D_Y$ to the right, so $(n-3)(n-4)/2$ matrices of the type~\eqref{eq:complexrotation2} drop out, leaving $n (n-1) - (n-3)(n-4) = 6 n -12$ real parameters in $V_R$, half of which are phases. Consequently we can restrict the product in Eq.~\eqref{eq:vr} to $\prod_{i=1}^3 \prod_{j=i+1}^n$.\footnote{In other words, the $U(n-3)$ subgroup of $U(n)$, acting on the lower right parts, can be modded out.} As expected, $\vec{Y}_\nu$ contains three (from $y_i$) plus six ($V_L$) plus $6 n -12$ ($V_R$) equals $6 n -3$ real, physical parameters, $3 (n-1)$ of which are phases. This agrees with the parameter counting in Ref.~\cite{Broncano:2002rw}.

The full mass matrix can then be written as
\begin{align}
 \M_\mathrm{full} = \matrixx{ 0 & v V_L^* D_Y^T V_R^T \\ v V_R D_Y V_L^\dagger & \M_R} ,
\label{eq:mfull}
\end{align}
with the explicit form for $V_L$:
\begin{align}
	V_L =
\matrixx{c_{12}^L c_{13}^L & s_{12}^L c_{13}^L & s_{13}^L e^{-i\beta_1^L}\\
	-c_{23}^L s_{12}^L- s_{23}^L s_{13}^L c_{12}^L e^{i\beta_1^L} & c_{23}^L c_{12}^L- s_{23}^L s_{13}^L s_{12}^L e^{i\beta_1^L} & s_{23}^L c_{13}^L\\
	s_{23}^L s_{12}^L- c_{23}^L s_{13}^L c_{12}^L e^{i\beta_1^L} & -s_{23}^L c_{12}^L- c_{23}^L s_{13}^L s_{12}^L e^{i\beta_1^L} & c_{23}^L c_{13}^L}  \,\diag (1, e^{i \beta_2^L/2}, e^{i\beta_3^L/2}) \,,
\label{eq:vl}
\end{align}
and the shorthand notation $s_{ij}^L = \sin \theta_{ij}^L$ etc. for the mixing angles. In the seesaw limit, we thus find
\begin{align}
 \M_\nu \simeq - v^2 \, V_L^* D_Y^T (V_R^\dagger \M_R V_R^*)^{-1} D_Y V_L^\dagger\,.
\label{eq:seesawvrvl}
\end{align}
Note that the occurring mass matrix $V_R^\dagger \M_R V_R^*$ is for $n>3$ no longer the most general complex symmetric $n\times n$ matrix, because we already restricted $V_R$ to the physical angles.

From Eq.~\eqref{eq:mfull} or Eq.~\eqref{eq:seesawvrvl} one can now extract the physical masses and mixing angles in terms of the parameters $y_i$, $\alpha_{ij}$, $\xi_{ij}$, $\theta_{ij}^L$ and $\beta_i^L$.

For completeness, let us count the total number of parameters in the lepton sector. We have $3$ charged lepton masses, $n$ heavy neutrino masses and $6 n -3$ real parameters in $\vec{Y}_\nu$, summing up to $7 n$ real parameters.

\section{Bottom-Up parametrization}
\label{sec:bottom-up}

Seeing as the seesaw formula~\eqref{eq:seesawvrvl} contains more parameters than measureable observables, it proves convenient to parametrize our ignorance of the high-energy sector of the seesaw model. In other words, there are infinitely many different $\vec{Y}_\nu$ and $\M_R$ that lead to the same low-energy neutrino parameters via the seesaw formula $\M_\nu \simeq - m_D^T \M_R^{-1} m_D$. The parametrization of this ambiguity is the goal of the Casas-Ibarra-parametrization~\cite{Casas:2001sr} of the Yukawa coupling:
\begin{align}
 \vec{Y}_\nu  = \frac{i}{v} \sqrt{\M_R}\, R \,\sqrt{d_m}\, U_\mathrm{PMNS}^\dagger \,.
\label{eq:cip}
\end{align}
This $\vec{Y}_\nu$ solves
\begin{align}
 -v^2 \, \vec{Y}_\nu^T \M_R^{-1} \vec{Y}_\nu \simeq \M_\nu = U_\mathrm{PMNS}^*\, d_m \,U_\mathrm{PMNS}^\dagger\,,
\end{align}
as long as the complex $n\times 3$ matrix $R$ satisfies $R^T R = \eins_{3\times 3}$. Correspondingly, any values of $\M_R$ and $R$ in Eq.~\eqref{eq:cip} lead to the same low-energy model, which means we can fix the PMNS mixing angles and masses to the observed ones and study $\vec{Y}_\nu$ independently in other processes.

We will now give an explicit parametrization of $R$, which has $3 n$ complex entries obeying $12$ real constraints ($R^T R = \eins$ is symmetric); we therefore expect to find $6 (n-2)$ real physical parameters. This agrees of course with the parameter counting in the last section, as we also have three charged lepton masses, three light neutrino masses, $n$ heavy neutrino masses and six parameters in the PMNS matrix, summing up to $7 n$ input parameters. Since we only need three orthonormal complex vectors $\in \mathbb{C}^n$ to define $R$, we can write 
\begin{align}
 R = O_{n\times n} \matrixx{\eins_{3\times 3}\\ \arrowvec{0}_3\\\vdots\\\arrowvec{0}_3} \equiv O_{n\times n} \eins_{n\times 3}\,,
\label{eq:svdR}
\end{align}
with a complex orthogonal $n\times n$ matrix $O_{n\times n} \in O(n,\mathbb{C})$. Similar to the $U(n)$ parametrization used for $V_R$ in the previous section, we can write $O_{n\times n}$ as a product of the $\Omega_{ij} (\gamma_{ij},\xi_{ij}')$ matrices from Eq.~\eqref{eq:complexrotation2}, but with the arguments $\xi_{ij}'=0$ and $\gamma_{ij} \in \mathbb{C}$ in order for $O_{n\times n}$ to be complex orthogonal:
\begin{align}
 O_{n\times n} = S\times \prod_{i=1}^n \prod_{j=i+1}^n \Omega_{ij} (\gamma_{ij},0) \,.
\label{eq:or}
\end{align}
Here, $S$ is a diagonal matrix with entries $\pm 1$, which decides the determinant of $O_{n\times n}$ and therefore whether it describes a rotation or a reflection.
We can once again mod out the $O(n-3,\mathbb{C})$ subgroup that acts on the zeros in $\eins_{n\times 3}$ by reordering the rotations, so the product in Eq.~\eqref{eq:or} can be restricted to $\prod_{i=1}^3 \prod_{j=i+1}^n$. Generally speaking, $R$ is an element of the quotient space $O(n,\mathbb{C})/O(n-3,\mathbb{C})$. For real angles $\gamma_{ij}$, $R$ is an element of the compact Stiefel manifold $V_3 (\mathbb{R}^n) \cong O(n,\mathbb{R})/O(n-3,\mathbb{R})$, which admits a Haar measure and could thus be used for statistical considerations.

Note that the case $n=4$ is special in the sense that there are no superfluous rotations that need to be modded out. One can therefore take any complete parametrization for $O_{4\times 4}$ without having to worry about unphysical parameters.

For complex angles, $O_{n\times n}$ and therefore $R$ and $\vec{Y}_\nu$ can have arbitrarily large elements, because the entries $\cos i x = \cosh x$ grow exponentially with $x$. As a result, the Yukawa couplings $\vec{Y}_\nu$ can be very large even for small $\M_R$, which is somewhat counterintuitive to the seesaw formula and relies on cancellations in the matrix product $\vec{Y}_\nu^T \M_R^{-1} \vec{Y}_\nu$. Even though these cancellations suggest finetuning of parameters, this is a viable possibility and deserves discussion, especially because a low $\M_R$ with large Yukawa couplings makes an LHC discovery (direct or indirect) possible~\cite{GeVseesaw}. Popular models to obtain this finetuning in a natural way are for example inverse~\cite{inverseseesaw} and linear seesaw~\cite{linearseesaw}, usually based on the case $n=6$ with an imposed structure in $\M_R$ and $\vec{Y}_\nu$ by some symmetry.

A useful constraint on the complex mixing angles $\gamma_{ij}$ comes from the perturbativity of $\vec{Y}_\nu$, on which many calculations involving $\vec{Y}_\nu$ rely. For this, we propose yet another, slightly different parametrization of $R$, which we obtain by noting that $\Omega_{ij} (\gamma_{ij},0) = \Omega_{ij} (\re (\gamma_{ij}),0) \times \Omega_{ij} (i \im(\gamma_{ij}),0)$, and that the ordering of the $\Omega_{ij}$ is arbitrary:
\begin{align}
 R = S\times \left(\prod_{i=1}^3 \prod_{j=i+1}^n \Omega_{ij} (\eta_{ij},0)\right)\, \left(\prod_{i=1}^3 \prod_{j=i+1}^n \Omega_{ij} (i \rho_{ij},0)\right) \eins_{n\times 3}\,.
\label{eq:svdR2}
\end{align}
The $6 (n-2)$ parameters that make up $R$ are thus equally divided among the new real parameters $\eta_{ij}$ and $\rho_{ij}$. The product involving $\eta_{ij}$ is just an orthogonal matrix, so we can restrict $\eta_{ij} \in [0,2\pi)$ or $[-\pi,\pi)$.\footnote{The parameter space of the angles can be further reduced using discrete field redefinitions like in Refs.~\cite{parameterspace,blennow}, which however lies outside the scope of this paper.} 

The parametrization~\eqref{eq:svdR2} can be more useful than~\eqref{eq:svdR},~\eqref{eq:or} because the $\eta_{ij}$ drop out of the expression $R^\dagger R$, which arises from (the often occurring) $\vec{Y}_\nu^\dagger \vec{Y}_\nu$ for degenerate $\M_R$.

\subsection{Perturbativity}
\label{sec:perturbativity}

Perturbativity of the Yukawa couplings gives constraints of the form~\cite{Casas:2010wm}
\begin{align}
 \tr (\vec{Y}_\nu^\dagger \vec{Y}_\nu)  = \frac{1}{v^2} \sum_{i=1}^n \sum_{j=1}^3 |R_{ij}|^2 M_i m_j \lesssim \mathcal{O}(1)\,.
\end{align}
This inequality is satisfied if 
\begin{align}
 |R_{ij}| \lesssim \frac{v}{\sqrt{3 n\, M_i m_j}} \simeq  10^6 \sqrt{\frac{3}{n}} \sqrt{\frac{\unit[1]{TeV}}{M_i}} \sqrt{\frac{\unit[10^{-2}]{eV}}{m_j}} \,,
\label{eq:perturbativitybound}
\end{align}
which constrains each entry in $R$. Note that the upper bound goes down for increasing $n$. Once the smallest masses are specified, the above equation can be used to give a total upper bound on the parameters $\rho_{ij}$. For example, the largest entry in $R$ can be estimated as $\cosh \rho_\mathrm{max}$ if just one $\rho_{ij}$ is large, and $(\cosh \rho_\mathrm{max})^{3 (n-2)}$ if all $\rho_{ij}$ are large. This typically leads to upper bounds on $|\rho_{ij}|$ of $\mathcal{O}(1$--$10)$.

\subsection{Lepton Flavor Violation}
\label{sec:lfv}

In the seesaw limit, the neutrino states $\nu_L$ that couple to the charged current are approximately given by~\cite{Grimus:2000vj} $\nu_L \simeq (\eins - \frac{1}{2} m_D^\dagger \M_R^{-2} m_D) U_\mathrm{PMNS} \nu_m + m_D^\dagger \M_R^{-1} N^c$, with the mass eigenstates $\nu_m$ and $N^c$. This leads to lepton flavor violating decays $\ell \rightarrow \ell' \gamma$~\cite{Cheng:1980tp}, of which $\mu \rightarrow e \gamma$ gives the strongest constraint:
\begin{align}
 \BR (\mu \rightarrow e \gamma ) \simeq \frac{3\alpha}{8\pi} \left| \left(m_D^\dagger \M_R^{-1} \,A \,\M_R^{-1} m_D\right)_{\mu e} -\frac{1}{2} \left( m_D^\dagger \M_R^{-2} m_D \,U_\mathrm{PMNS}\, B\,  U_\mathrm{PMNS}^\dagger  + \hc \right)_{\mu e} \right|^2 ,
\label{eq:branchingratiomuegamma}
\end{align}
with the diagonal matrices $A_{ij} \equiv \delta_{ij} \, g(M_j^2/M_W^2)$ and $B_{ij} \equiv \delta_{ij}\,  g(m_j^2/M_W^2)$ and the monotonic decreasing loop function
\begin{align}
 g (x) = \frac{10 - 43 x + 78 x^2 - 49 x^3 + 4 x^4 + 18 x^3 \log x}{6 (x-1)^4} \,, && g(0) = \frac{5}{3}\,, && g(1) = \frac{17}{12}\,, && g(x\rightarrow \infty) = \frac{2}{3} \,.
\end{align}
Here we already omitted the contribution from the light neutrinos, as the corresponding term $(U_\mathrm{PMNS} \,B\, U_\mathrm{PMNS}^\dagger)_{\mu e}$ is suppressed by the tiny mass-squared differences of the active neutrinos~\cite{Cheng:1980tp}. The remaining contributions are naively of order $(m_D/\M_R)^2$, but can be enhanced via $R$. In the approximation $M_j \gg M_W \gg m_k$ we have $A\propto \eins \propto B$, so Eq.~\eqref{eq:branchingratiomuegamma} simplifies to
\begin{align}
  \BR (\mu \rightarrow e \gamma ) \simeq \frac{3\alpha}{8\pi} \left|\left( m_D^\dagger \M_R^{-2} m_D \right)_{\mu e} \right|^2 \simeq \frac{3\alpha}{8\pi} \left|\left( U_\mathrm{PMNS} \sqrt{d_m} R^\dagger \M_R^{-1} R \sqrt{d_m} U_\mathrm{PMNS}^\dagger \right)_{\mu e} \right|^2 .
\label{eq:mutoegamma}
\end{align}
Since this branching ratio can easily exceed the current limit $\BR (\mu \rightarrow e \gamma ) < 2.4\times 10^{-12}$~\cite{meg}---for perturbative couplings---it gives constraints on the $\rho_{ij}$. Naively, the relevant product in Eq.~\eqref{eq:mutoegamma} is $|R|^2 m/M$, so the LFV constraints have a different structure than the perturbativity bound~\eqref{eq:perturbativitybound}. Note that this constraint is connected to the unitarity of the lepton mixing matrix~\cite{Antusch:2006vwa}, because the relevant matrix $(\eins - \frac{1}{2} m_D^\dagger \M_R^{-2} m_D) U_\mathrm{PMNS}$ is no longer unitary at this order. Consequently, the above bound also ensures the validity of our parametrization for $\vec{Y}_\nu$ from Eq.~\eqref{eq:cip}, as the matrix $U_\mathrm{PMNS}$ in $\vec{Y}_\nu$ only corresponds to the lepton mixing matrix if $(m_D^\dagger \M_R^{-2} m_D)_{ij}\ll 1$.

While not particularly nice to look at analytically, the constraints given here should be useful in numerical scans involving $\vec{Y}_\nu$ for arbitrary many right-handed neutrinos, extending for example the analysis of Ref.~\cite{Casas:2010wm}.

\section{Application to Neutrino Anarchy}
\label{sec:anarchy}

The idea that the large lepton mixing angles and small neutrino hierarchy are not due to some flavor symmetry but rather the absence of any distinction between the neutrino generations has been around for over a decade~\cite{anarchy,Haba:2000be}. The basis independence of the neutrino mass and mixing matrices leads to a distribution of the mixing angles according to the Haar measure of the Lie group $G$ that diagonalizes $\M_\nu$ (so $G= U(3)$ for complex, $G= O(3)$ for real $\M_\nu$). Since these distributions prefer large mixing angles, this ansatz seems to fit well to the observed large leptonic mixing angles (see also, however, a critique of the anarchy approach in Refs.~\cite{critic1,critic2}). While the distribution of the mixing angles is uniquely given by the Haar measure, the distributions of the masses and Yukawa couplings are not unique, and usually the simplest (linear) measure is used (listed in App.~\ref{sec:measures}).

\begin{figure}[tb]
\setlength{\abovecaptionskip}{-2ex}
	\begin{center}
		\includegraphics[width=0.46\textwidth]{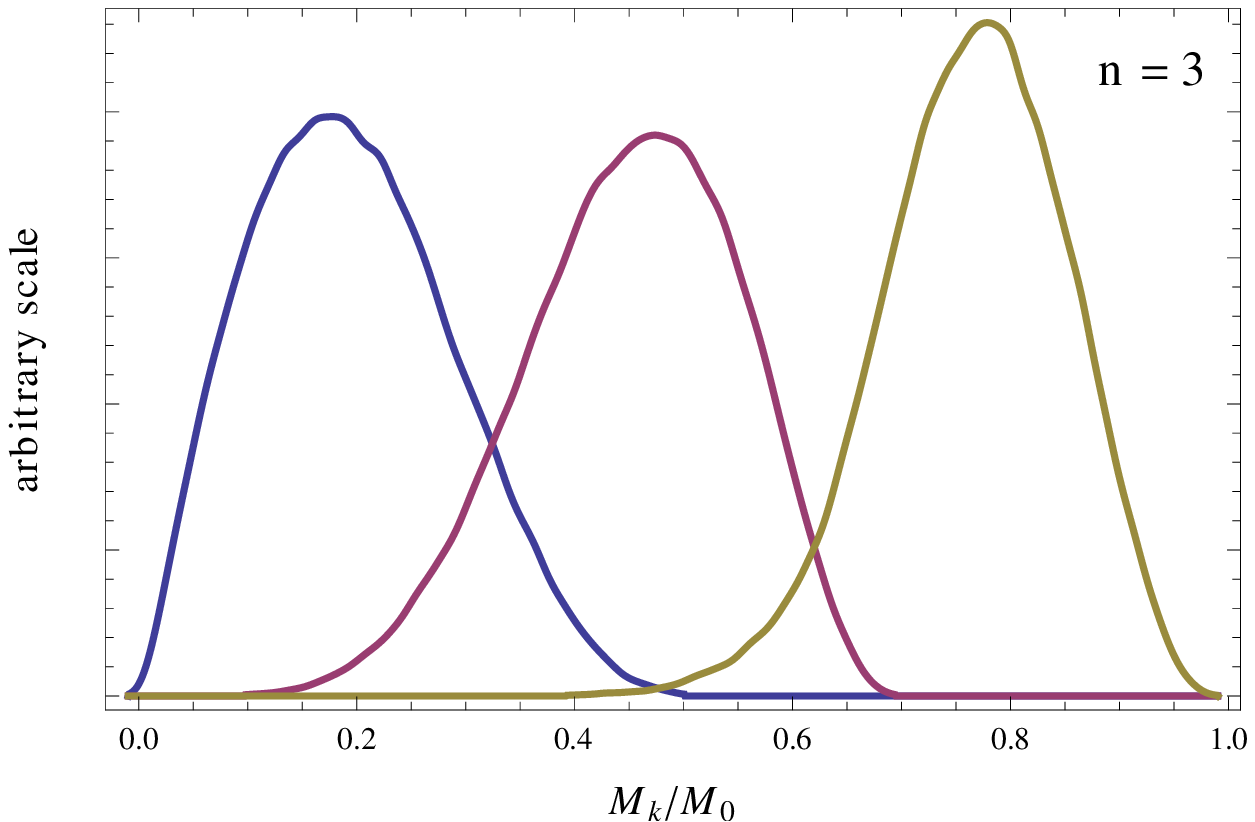}
		\includegraphics[width=0.46\textwidth]{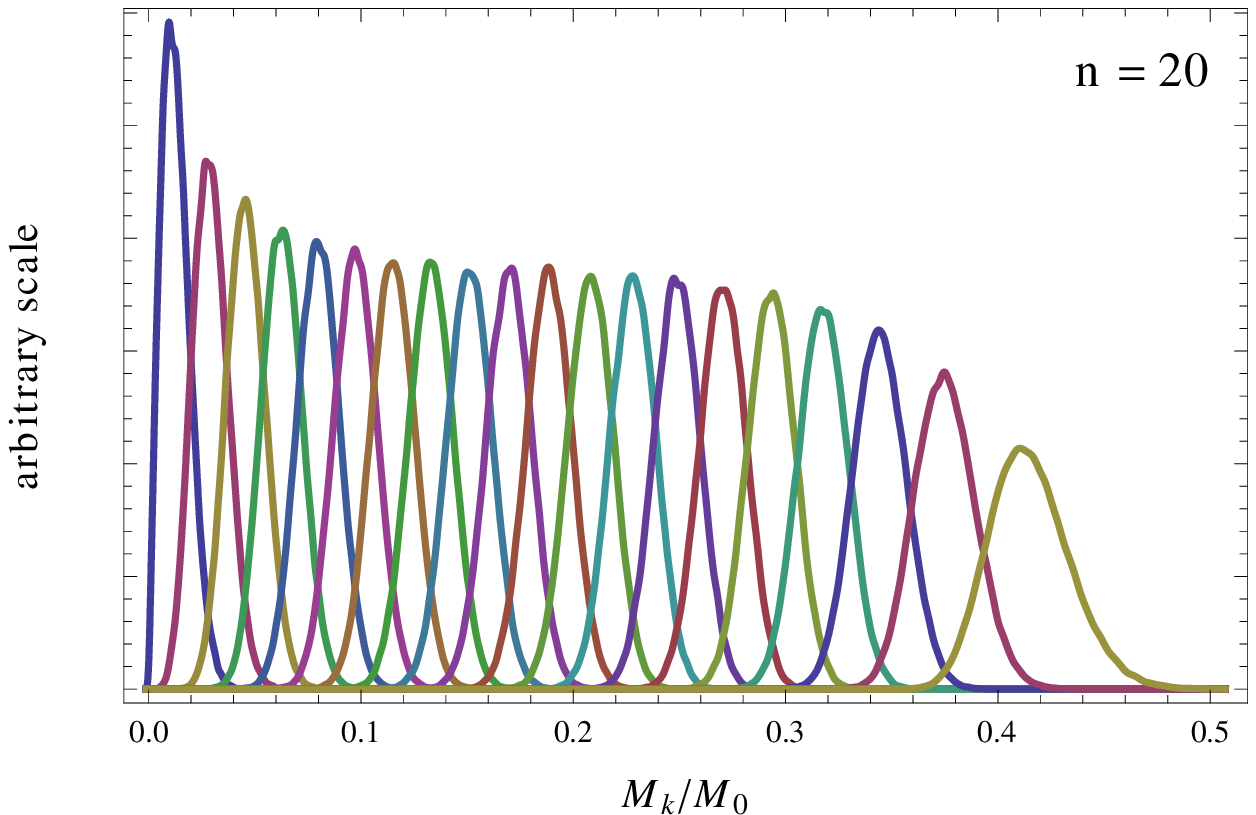}
	\end{center}
		\caption{Distribution of the $\M_R$ eigenvalues $M_k$ according to the linear measure from Eq.~\eqref{eq:linearmajoranameasure} with the boundary constraint $\sum_{k=1}^n M_k^2 \leq M_0^2$.}
	\label{fig:massdist}
\end{figure}

\begin{figure}[tb]
\setlength{\abovecaptionskip}{-2ex}
	\begin{center}
		\includegraphics[width=0.5\textwidth]{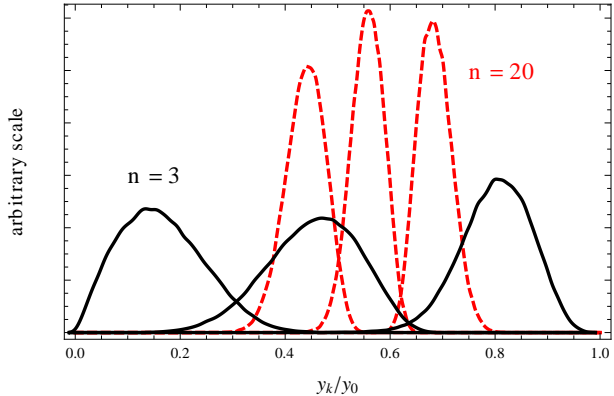}
	\end{center}
		\caption{Distribution of the singular values $y_k$ of the Yukawa matrix according to the linear measure from Eq.~\eqref{eq:linearyukawameasure} for $n=3$ (black) and $n=20$ (red/dashed) with the boundary constraint $\sum_{k=1}^3 y_k^2 \leq y_0^2$.}
	\label{fig:yukawadist}
\end{figure}

Let us now discuss neutrino anarchy, i.e.~basis independence, with $n$ right-handed neutrinos.
In the seesaw limit we have the low-energy mass matrix for the active neutrinos from Eq.~\eqref{eq:seesawvrvl}
\begin{align}
 \M_\nu \simeq - v^2 \, V_L^* D_Y^T V_R^T \M_R^{-1} V_R D_Y V_L^\dagger\,,
\label{eq:anarchy_seesaw}
\end{align}
where $V_L$ and $V_R$ are now assumed to be distributed according to the Haar measure of $U(3)$ and $U(n)$, respectively, while we take the simplest linear measure for the eigenvalues of $\M_R$~\cite{Haba:2000be}
\begin{align}
 \dd \M_R \propto \prod_{i<j}^n |M_j^2 - M_i^2| \prod_{k=1}^n M_k \,\dd M_k\,.
\label{eq:linearmajoranameasure}
\end{align}
We employ the invariant boundary $\tr ( \M_R^\dagger \M_R )= \sum M_i^2 \leq M_0^2$ for the scanning region; see Fig.~\ref{fig:massdist} for a visualization of the distribution.
For the singular values of $D_Y$ we use the linear measure (see App.~\ref{sec:measures} for the derivation)
\begin{align}
 \dd D_Y \propto  \prod_{m=1}^3 (y_m^2)^{n-3}\prod_{i<j}^3 (y_i^2-y_j^2)^2 \prod_{k=1}^3 y_k \dd y_k\,,
\label{eq:linearyukawameasure}
\end{align}
with a similar boundary $\tr ( \vec{Y}_\nu^\dagger \vec{Y}_\nu ) = \tr (D_Y^\dagger D_Y)= \sum_j^3 y_j^2 \leq y_0^2$. The effect of large $n$ will be a reduced hierarchy in the singular values $y_i$ (see Fig.~\ref{fig:yukawadist}). Note that Eq.~\eqref{eq:linearyukawameasure} is only valid for $n\geq 3$, see App.~\ref{sec:nequal2} for the used measure for $n=2$.

We will now briefly describe how the random matrices in this paper are generated. The random unitary matrices $V_L$ and $V_R$ in Eq.~\eqref{eq:anarchy_seesaw}---following the Haar distribution---can efficiently be obtained from a QR decomposition, as described in Refs.~\cite{random_matrix}. Drawing eigenvalues from Eq.~\eqref{eq:linearmajoranameasure} is a bit more complicated. The procedure usually employed in the literature is the creation of a symmetric matrix $\M_R$, where $\re (\M_R)_{ij}$ and $\im (\M_R)_{ij}$ are uniformly distributed in $[-M_0,\ M_0]$. To stay rotationally invariant, only matrices with $\tr ( \M_R^\dagger \M_R )\leq M_0^2$ are then used and diagonalized (roughly speaking, we change the scanning region from a hypercube to a hypersphere). This method works well for $n\leq 3$, but becomes virtually unusable for $n\geq 4$, because the volume of the hypercube increases sharply with $n$ compared to the hypersphere, making it hard to accumulate statistics. For this reason, we draw the eigenvalues directly out of the probability distribution~\eqref{eq:linearmajoranameasure} using a multivariate Metropolis algorithm~\cite{metropolis}. This Markov chain Monte Carlo method is sufficiently fast even for large $n$ and conveniently works without knowing the normalization factor of the distribution. We use a multivariate Gaussian proposal distribution to generate the next steps in the chain and pick only every hundredth (tenth for $n=40$) valid point to reduce correlations in the samples.

The value $v^2 y_0^2/M_0$ will fix the overall light neutrino mass scale and can be used to fix one of the mass-squared differences $\Delta m_{ij}^2$.
The PMNS mixing angles and phases do not change with $n$ and are given by the Haar measure for $U(3)$~\cite{Haba:2000be} (ignoring unphysical phases):
\begin{align}
 \dd U_\mathrm{PMNS} \propto \dd s_{12}^2\, \dd s_{23}^2\, \dd c_{13}^4\, \dd \delta\, \dd \alpha\, \dd \beta\,,
\end{align}
so the only effect of $n\neq 3$ is a change of the three eigenvalues of $\M_\nu$, i.e.~the ratio $R_\nu \equiv (m_2^2-m_1^2)/(m_3^2-m_1^2)$, where we sorted the masses like $m_1 \leq m_2 \leq m_3$.
In this notation, $R_\nu<1/2$ corresponds to normal hierarchy (NH) and $R_\nu>1/2$ to inverted hierarchy (IH), with best-fit values $R_\nu^\mathrm{NH} = 0.03$ and $R_\nu^\mathrm{IH} =0.97$, respectively, taken from a recent global fit~\cite{Tortola:2012te}.

\begin{figure}[tb]
\setlength{\abovecaptionskip}{-2ex}
	\begin{center}
		\includegraphics[width=0.98\textwidth]{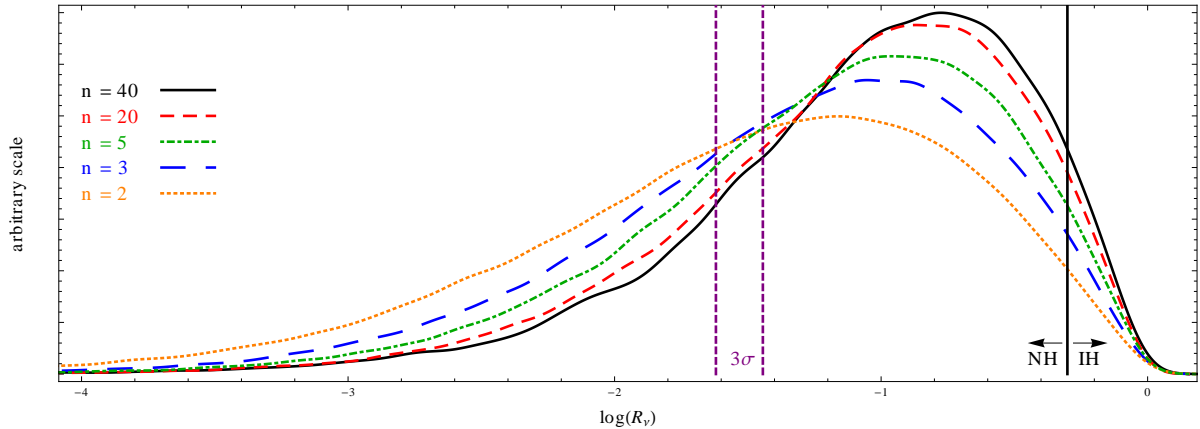}
	\end{center}
		\caption{$\log_{10}[R_\nu] \equiv \log_{10}\left[(m_2^2-m_1^2)/(m_3^2-m_1^2)\right]$ for various $n$. The purple/dashed vertical lines indicate the $3\sigma$ range for $R_\nu\simeq 0.03$~\cite{Tortola:2012te} (for NH). The black vertical line divides the NH and IH solutions.}
	\label{fig:ratios}
\end{figure}

We observe from Fig.~\ref{fig:ratios} that for increasing $n$, the distribution for $R_\nu$ shifts to larger values, while the width decreases. This reduction of hierarchy can be tracked back mainly to the Yukawa couplings, as the factor $\prod_{m=1}^3 (y_m^2)^{n-3}$ in $\dd D_Y$ pulls the $y_i$ tightly together. An analysis without this factor shows that the behavior of $R_\nu$ goes in the opposite direction, i.e.~the maximum shifts toward small values for large $n$ while being diluted. However, the distribution then quickly converges as the only $n$-dependend change comes from $V_R^T \M_R^{-1} V_R$, for which only the first dozen or so $M_i$ are relevant, due to the hierarchy in $\M_R$. Correspondingly, to shift the maximum of $R_\nu$ to its observed value, one has not only to omit the factor $\prod_{m=1}^3 (y_m^2)^{n-3}$ in $\dd D_Y$, but also to increase the hierarchy by inserting something like $\prod_{i<j}(y_i^2-y_j^2)^2$, i.e.~using a non-linear measure (see Ref.~\cite{Feldstein:2011ck} for other $y_i$ distributions that accomplish this task). We conclude that the large $n$ limit in the anarchy approach worsens the agreement with data.
While the one data point nature provides can obviously not be used to find the overlying distribution, it should be fair to say that anarchy works best for small $n$.

Let us finally comment on the limit $n\rightarrow \infty$. As the Yukawa couplings $y_i$ become quasi-degenerate for very large $n$, the neutrino masses are simply given by the upper left $3\times 3$ submatrix of $V_R^T \M_R^{-1} V_R$ (times a prefactor). Since there is no preferred direction in the $V_R$ rotations, the entries $(V_R)_{ij}$ have the same mean magnitude for a given $n$. The mean entries $\langle (V_R^T \M_R^{-1} V_R)_{jk}\rangle$ then take the form of a random walk with decreasing step-size $\sum_m^n e^{i \theta_{m}}/M_m$, and due to the hierarchy in the singlet masses $M_k$, only the first couple of steps are relevant. The mean magnitudes are again the same for all entries, so the only structure in the neutrino mass matrix comes from random phases $\tilde \theta_{j k}$: $\langle (V_R^T \M_R^{-1} V_R)_{jk}\rangle \propto e^{i \tilde\theta_{jk}}$. So, if the number $n$ is high enough, anarchy eventually leads to democracy. It is easy to show that a matrix of this type predominantly yields NH solutions, so even though IH becomes more and more probable for large $n$, it will never dominate.

\section{Conclusion}

One of the simplest ways to explain the masses of the active neutrinos is the introduction of right-handed partners. As the number $n$ of these SM singlets is in principle unconstrained, it is interesting to study implications of varying $n$.
We have given explicit parametrizations for the mixing and coupling matrices connecting the $n$ right-handed neutrinos to the SM in terms of physical parameters, both in the top-down and the bottom-up approach (Casas-Ibarra-parametrization). For the latter, constraints on the involved parameters from perturbativity of the Yukawa couplings as well as charged lepton flavor violation have been discussed.

As a novel application of the $n$ singlet framework we studied basis independence in the neutrino sector, i.e.~anarchy. Of the low-energy neutrino parameters, only the neutrino mass distribution changes with $n$, and we showed that anarchy---with the simplest linear measure---seems to prefer small $n$ in view of the observed mass-squared differences.

\begin{acknowledgments}
The author thanks Werner Rodejohann and Mattias Blennow for reading the manuscript and offering valuable comments.
This work was supported by the ERC under the Starting Grant MANITOP, by the Max Planck Society through the Strategic Innovation Fund, and by the IMPRS-PTFS.
\end{acknowledgments}

\appendix

\section{The Group \texorpdfstring{$U(N)$}{U(N)}}
\label{sec:thegroupUN}

For convenience we provide a short and rather colloquial review of the group $U(N,\mathbb{C}) \equiv U(N)$. Formally one can define it as the set of complex $N\times N$ matrices that leave the inner product of $\mathbb{C}^N$ invariant, i.e.~$U(N) = \{V \in \mathbb{C}^{N\times N}\, |\, V^\dagger V = \eins\}$. With a little effort one can further show that the entries of $U(N)$ form a group under matrix multiplication, actually even a compact connected Lie group. Counting the number of entries in $V\in U(N)$ and constraints, we find that $U(N)$ has $N^2$ real parameters. One can show that every element in $U(N)$ can be written as $V = \exp ( A )$ with skew-hermitian $A = - A^\dagger$. We can choose a basis for $A = \sum \omega_j X_j$ such that the $N^2$ linearly independent $X_j$ satisfy Lie algebra relations $[X_k, X_l] = i f_{klm} X_m$ with structure constants $f_{klm}$.

The $N^2$ generators $X_j$ can be further separated into $N (N+1)/2$ symmetric generators $X_s = i Y = i Y^T$ and $N (N-1)/2$ antisymmetric real generators $X_a = -X_a^T$. It also proves convenient to treat the $N$ diagonal (symmetric) generators $Y_d$ separately, as they are especially easy to exponentiate. To fix the normalization, we chose $(Y_d^k)_{ij} = \delta_{ij} \delta_{jk}$, while the $N (N-1)/2$ non-diagonal $Y_s$ and $X_a$ have just two non-vanishing entries, namely $+1$ in the upper right half and $\pm 1$ in the lower left (plus sign for $Y_s$, minus sign for $X_a$). Note that this normalization---and therefore also $f_{klm}$---differs from the common convention, but is of no importance in the following (see also Refs.~\cite{arbitraryn}). A general element of $U(N)$ can then be written as the product of all rotations:
\begin{align}
 U(N) \ni V = \prod_{j=1}^N \exp (i \alpha_j Y_d^j) \prod_{k=1}^{N (N-1)/2}  \exp (i \beta_k Y_s^k)  \prod_{m=1}^{N (N-1)/2}  \exp (\gamma_m X_a^m) \,.
\label{eq:generalun}
\end{align}
The equivalence of this parametrization to the initial $V = \exp ( A )$ can be proven using the Baker-Campbell-Hausdorff formula and the fact that the generators satisfy a Lie algebra.
The order of the rotations in Eq.~\eqref{eq:generalun} is of course arbitrary~\cite{Gronau:1985kx} and should be chosen to simplify given expressions. The three different types of rotations take the form
\begin{align}
e^{i \beta Y_s} &= \matrixx{ 1 & &  & & & \\ & \ddots & & & &  & \\ & & \cos (\beta) & &  i \sin (\beta) &  & \\ & & & \ddots & &  & \\ & & i \sin (\beta) & & \cos (\beta) &  & \\ & & & & &\ddots  & \\ & & & & & & 1\\} ,&&
e^{\gamma X_a} &= \matrixx{ 1 & &  & & & \\ & \ddots & & & &  & \\ & & \cos (\gamma) & &  \sin (\gamma) &  & \\ & & & \ddots & &  & \\ & & - \sin (\gamma) & & \cos (\gamma) &  & \\ & & & & &\ddots  & \\ & & & & & & 1\\} ,
\end{align}
and
\begin{align}
  \exp \left(i \alpha Y_d\right) &= \diag (1,\,\dots,\,1,\,\exp(i \alpha),\,1,\,\dots,\,1) \,,
\end{align}
with real angles $\alpha,\beta,\gamma$. 
Another useful parametrization for a general unitary matrix along the same lines as above is given by
\begin{align}
 U(N) \ni V = \prod_{j=1}^N \exp (i \alpha_j' Y_d^j) \prod_{k=1}^{N (N-1)/2}  \exp (i \beta_k' Y_s^k+\gamma_k' X_a^k)  \,,
\label{eq:generalun2}
\end{align}
where we combined the generators $X_a$ and $Y_s$ that have the same vanishing entries. This yields a product of matrices of the type
\begin{align}
  \exp \left(i \beta_j' Y_s^j+\gamma_j' X_a^j \right) = \matrixx{ 1 & &  & & & \\ & \ddots & & & &  & \\ & & \cos (z_j) & &  \sin (z_j) e^{i \xi_j} &  & \\ & & & \ddots & &  & \\ & & - \sin (z_j) e^{-i \xi_j} & & \cos (z_j) &  & \\ & & & & &\ddots  & \\ & & & & & & 1\\} ,
\end{align}
with the real parameters $z_j = |u_j|$, $\xi_j = \arg (u_j)$ and $u_j = i \beta_j' - \gamma_j'$ in our chosen normalization. Once again, the ordering of the matrices in Eq.~\eqref{eq:generalun2} is arbitrary.

\section{Measures for Anarchy}
\label{sec:measures}

This appendix provides the necessary measures for the anarchy approach in Sec.~\ref{sec:anarchy}. We refer to Ref.~\cite{Haba:2000be} for a detailed derivation of the known measures, and simply quote the result for real and complex Majorana $N\times N$ matrices:
\begin{align}
 \dd \M_\mathrm{real} \propto \prod_{i<j}^N |D_i-D_j| \prod_{k=1}^N \dd D_k \dd O\,, &&
 \dd \M_\mathrm{complex} \propto \prod_{i<j}^N |D_i^2-D_j^2| \prod_{k=1}^N D_k \dd D_k \dd U\,,
\label{eq:majoranameasure}
\end{align}
where the decompositions $\M_\mathrm{real} = O D O^T$ and $\M_\mathrm{complex} = U D U^T$ were used. $\dd O$ and $\dd U$ denote the usual Haar measure for $O(N)$ and $U(N)$, respectively, while the diagonal matrix $D$ contains the eigenvalues of $\M$. We will omit the calculation of the normalization factor for all measures in this appendix, as they are irrelevant for our considerations. It can in principle be calculated by integration over some volume, e.g.~$\tr (\M_\mathrm{complex}^\dagger \M_\mathrm{complex}) \leq M_0^2$, up to the arbitrary overall mass scale $M_0$.

For Dirac matrices, the measures read~\cite{Haba:2000be}
\begin{align}
 \dd m_\mathrm{real} \propto \prod_{i<j}^N |D_i^2-D_j^2| \prod_{k=1}^N \dd D_k \dd O_L \dd O_R\,, &&
 \dd m_\mathrm{complex} \propto \prod_{i<j}^N (D_i^2-D_j^2)^2 \prod_{k=1}^N D_k \dd D_k \dd U_L \dd U_R\,,
\label{eq:diracmeasure}
\end{align}
with $m_\mathrm{real} = O_R D O_L^T$, $m_\mathrm{complex} = U_R D U_L^\dagger$, and the Haar measures for $O(N)$ $\dd O_{L,R}$ and $U(N)$ $\dd U_{L,R}$ respectively. For $\dd m_\mathrm{complex}$, unphysical phases in the overlap of $\dd U_L$ and $\dd U_R$ should be modded out, once a parametrization is specified.

These invariant measures are of course not unique, but rather the simplest consistent ansatz that give the right mass dimension.\footnote{For example, $\M_\mathrm{complex}$ contains $N (N+1)$ independent real parameters, so the measure $\dd \M_{ij}$ has mass dimension $N (N+1)$. Rewriting $\dd \M_{ij}$ in the form~\eqref{eq:majoranameasure} matches this mass dimension, as can be easily verified.}
With the above equations, one can easily write down the measure for an hermitian matrix $H = U h U^\dagger = H^\dagger$ under $U(N)$:
\begin{align}
 \dd H \propto \prod_{i<j}^N (h_i - h_j)^2 \prod_{k=1}^N \dd h_k \dd U \,.
\label{eq:hermitianmeasure}
\end{align}
As a cross-check, one can plug the hermitian matrix $m_\mathrm{complex}^\dagger m_\mathrm{complex}$ into Eq.~\eqref{eq:hermitianmeasure} to rederive Eq.~\eqref{eq:diracmeasure}.
We can now give a measure for a non-quadratic $n\times 3$ complex Dirac matrix $M_{SVD} = U_R D_Y U_L^\dagger$ with $U_R \in U(n)$, $U_L \in U(3)$. Considering $M_{SVD}^\dagger M_{SVD}$ shows that $\dd M_{SVD}$ should at least contain a factor
\begin{align}
 \prod_{i<j}^3 (y_i^2-y_j^2)^2 \prod_{k=1}^3 y_k \dd y_k \,,
\end{align}
which however has too small of a mass dimension. The same procedure for the $n\times n$ hermitian $ M_{SVD} M_{SVD}^\dagger$ runs into problems, because $n-3$ of the eigenvalues are zero and would give a vanishing measure. Taking the product only over the non-vanishing factors however yields
\begin{align}
 \dd M_{SVD} \propto \prod_{m=1}^3 (y_m^2)^{n-3}\prod_{i<j}^3 (y_i^2-y_j^2)^2 \prod_{k=1}^3 y_k \dd y_k \dd U_L \dd U_R\,,
\label{eq:svdmeasure}
\end{align}
which has the right mass dimension ($6 n$) and can therefore be viewed as the linear measure for $M_{SVD}$. The overlap of $\dd U_L$ and $\dd U_R$---this time including angles, see Sec.~\ref{sec:top-down}---should again be modded out in a given parametrization to avoid overcounting of physically equivalent configurations (similar to gauge-fixing, as pointed out in Ref.~\cite{critic2}).

The measure~\eqref{eq:svdmeasure} can of course be easily extended to a general measure over $O(N)\times O(M)$ or $U(N)\times U(M)$ with $M\leq N$:
\begin{align}
 \dd M_{\mathrm{real}} \propto \prod_{i<j}^M |y_i^2-y_j^2| \prod_{k=1}^M y_k^{N-M} \dd y_k \dd O_L \dd O_R\,, &&
 \dd M_{\mathrm{complex}} \propto \prod_{i<j}^M (y_i^2-y_j^2)^2 \prod_{k=1}^M y_k^{1+2 (N-M)} \dd y_k \dd U_L \dd U_R\,,
\label{eq:generalsvdmeasure}
\end{align}
with the $N\times M$ matrices $M_\mathrm{real} = O_R D_Y O_L^T$ and $M_\mathrm{complex} = U_R D_Y U_L^\dagger$. Unphysical rotations need to be modded out again. A simple check shows that the mass dimensions are right and $N=M$ indeed leads back to Eq.~\eqref{eq:diracmeasure}. Eq.~\eqref{eq:generalsvdmeasure} can actually be obtained from exactly these requirements, without the need to use hermitian matrices in the derivation. For example, $\dd M_\mathrm{complex}$ should contain the known $N=M$ factor $\prod_{i<j}^M (y_i^2-y_j^2)^2 \prod_{k=1}^M \dd y_k$, which leaves some factor of mass dimension $2 M$ to the power of $N-M$. Since a factor $\prod_{i<j}^M (y_i^k - y_j^k)^m$---which would increase the hierarchy in the $y_i$---has mass dimension $M (M-1) k m/2$, it can only be used in special cases, not for general $M$ (at least for integer $k$ and $m$). Thus, for general $M$, the only factor with mass dimension $2 M$ is $\prod^M_i y_i^2$, leading back to Eq.~\eqref{eq:generalsvdmeasure}.

\section{The Case \texorpdfstring{$n=2$}{n=2}}
\label{sec:nequal2}

In this appendix we collect the formulas for the case $n=2$ (see also Refs.~\cite{twosinglets,n=2}), which is a little different than the $n\geq 3$ cases and would interrupt the flow if included in the main text. With the $2\times 2$ diagonal matrix $\M_R$ we find the top-down parametrization in the singular value decomposition for the Yukawa coupling
\begin{align}
 \vec{Y}_\nu = V_R \matrixx{y_1 & 0 & 0\\ 0 & y_2 & 0} V_L^\dagger \,.
\end{align}
Two phases of $V_R$ can be absorbed by $V_L^\dagger$, three of $V_L$ in the lepton fields. Since one of the diagonal phase rotations of $V_L$ acts on the zeros, there are only two phases in $V_L$ left, so we can write
\begin{align}
 V_R &= \matrixx{\cos (\alpha_{12} ) & \sin (\alpha_{12}) e^{i \xi_{12}} \\ - \sin ( \alpha_{12}) e^{-i \xi_{12}} & \cos ( \alpha_{12}) } , \\
V_L &=
\matrixx{c_{12}^L c_{13}^L & s_{12}^L c_{13}^L & s_{13}^L e^{-i\beta_1^L}\\
	-c_{23}^L s_{12}^L- s_{23}^L s_{13}^L c_{12}^L e^{i\beta_1^L} & c_{23}^L c_{12}^L- s_{23}^L s_{13}^L s_{12}^L e^{i\beta_1^L} & s_{23}^L c_{13}^L\\
	s_{23}^L s_{12}^L- c_{23}^L s_{13}^L c_{12}^L e^{i\beta_1^L} & -s_{23}^L c_{12}^L- c_{23}^L s_{13}^L s_{12}^L e^{i\beta_1^L} & c_{23}^L c_{13}^L}  \,\diag (1, e^{i \beta_2^L/2}, 1) \,,
\end{align}
with the shorthand notation $s_{ij}^L = \sin \theta_{ij}^L$ etc. for the mixing angles. The linear measure for $y_{1,2}$ for the anarchy framework can be derived along the same lines as in App.~\ref{sec:measures}, with the result
\begin{align}
 \dd D_Y \propto (y_1^2 - y_2^2)^2 y_1^3 y_2^3 \dd y_1 \dd y_2\,.
\end{align}
Finally, the bottom-up parametrization is given by Eq.~\eqref{eq:cip} with two different $R$ matrices depending on the low-energy neutrino hierarchy ($\gamma \in \mathbb{C}$)~\cite{Ibarra:2003up}:
\begin{align}
 R_\mathrm{NH} = \matrixx{0 & \cos \gamma & - \sin \gamma \\ 0 & \sin \gamma & \cos \gamma} , &&
 R_\mathrm{IH} = \matrixx{\cos \gamma & - \sin \gamma & 0\\ \sin \gamma & \cos \gamma & 0} .
\end{align}


\end{document}